\documentclass[final]{aipproc}

\layoutstyle{6x9}

\usepackage{amsmath}

\DeclareMathOperator{\Tr}{Tr}
\newcommand{\xbj}{x}
\newcommand{\slim}{\mskip 1.5mu}
\newcommand{\bm}{\mathbf}
\newcommand{\PhiA}{\tilde{\Phi}_{A}}
\newcommand{\DeltaA}{\tilde{\Delta}_{A}}
\newcommand{\cdott}{{\mskip -1.5mu} \cdot {\mskip -1.5mu}}

\begin{document}

\title{Transverse-Momentum Dependent Functions in Semi-Inclusive DIS}

%\classification{}

\author{Alessandro Bacchetta}{
address={Theory Group, DESY,
22603 Hamburg, Germany}
}

\begin{abstract}
The cross section for semi-inclusive deep inelastic scattering can be
decomposed in terms of 18 structure functions. At low transverse momentum of
the detected hadron, the structure functions can be expressed in terms of
transverse-momentum-dependent 
parton distribution and fragmentation functions. Here, a few selected 
examples are
illustrated and discussed.
\end{abstract}

\maketitle

I present a selection of results from a recent work~\cite{Bacchetta:2006tn}, where semi-inclusive deep
inelastic scattering, 
$
  \label{sidis}
\ell(l) + N(P) \to \ell(l-q) + h(P_h) + X
$,
%($\ell$ denotes the beam lepton, $N$ the nucleon target, $h$
%the produced hadron), 
was analyzed in the regime of low transverse momentum of the detected
hadron,  with the
goal
of completing the existing literature on the subject. 
In the following, the standard variables 
$Q^2$, $\xbj$, $y$, $z$ are used. $M$ and $M_h$ denote the respective
masses of the nucleon and of the hadron $h$.  
Azimuthal angles are defined in accordance to the 
Trento conventions~\cite{Bacchetta:2004jz}. $P_{h\perp}$ denotes the component
of $P_h$ perpendicular to $q$, in, e.g., the proton-photon Breit frame.

Assuming single photon exchange, the lepton-hadron cross section can be
computed as the contraction of the hadronic and the leptonic tensor
and expressed by a set of structure
functions. For unpolarized
beam and target, the cross section contains only four structure functions.  
Neglecting corrections of order $1/Q^2$, it can be written as
%\vspace{-3pt}
\begin{equation}
\begin{split}
 \frac{d\sigma}{d \Gamma} &=  \frac{2 \pi\,\alpha^2\slim y}{8 z\slim Q^4}\, 2 M W^{\mu \nu}\,
  L_{\mu \nu} = \frac{2 \pi \alpha^2}{\xbj y\slim Q^2}\,
\Bigl\{
\left(1-y +y^2/2\right)\,F_{UU ,T} +
(1-y)\;F_{UU ,L}
\\ & \quad
+(2-y)\, \sqrt{1-y}\,\cos\phi_h\;F_{UU}^{\cos\phi_h}
+(1-y)\,\cos(2\phi_h)\;F_{UU}^{\cos 2\phi_h}\Bigr\},
\end{split} 
%\vspace{-3pt}
\end{equation} 
where $d \Gamma = dx\,dy\,dz\,d\phi_h\,dP_{h\perp}^2$. 
The structure functions depend on $\xbj$, $Q^2$, $z$ and
$P_{h\perp}^2$. There is no established notation for the 
structure functions.  Here as in \cite{Bacchetta:2006tn}, 
the superscript indicates the azimuthal modulation 
generated in the cross section. The first and
second subscript indicate the
respective polarization of beam and target. When needed, 
the third subscript specifies the polarization of
the virtual photon.

\begin{figure}[ht]
%\begin{center}
\begin{tabular}{ccccc}
\includegraphics[height=3.5cm]{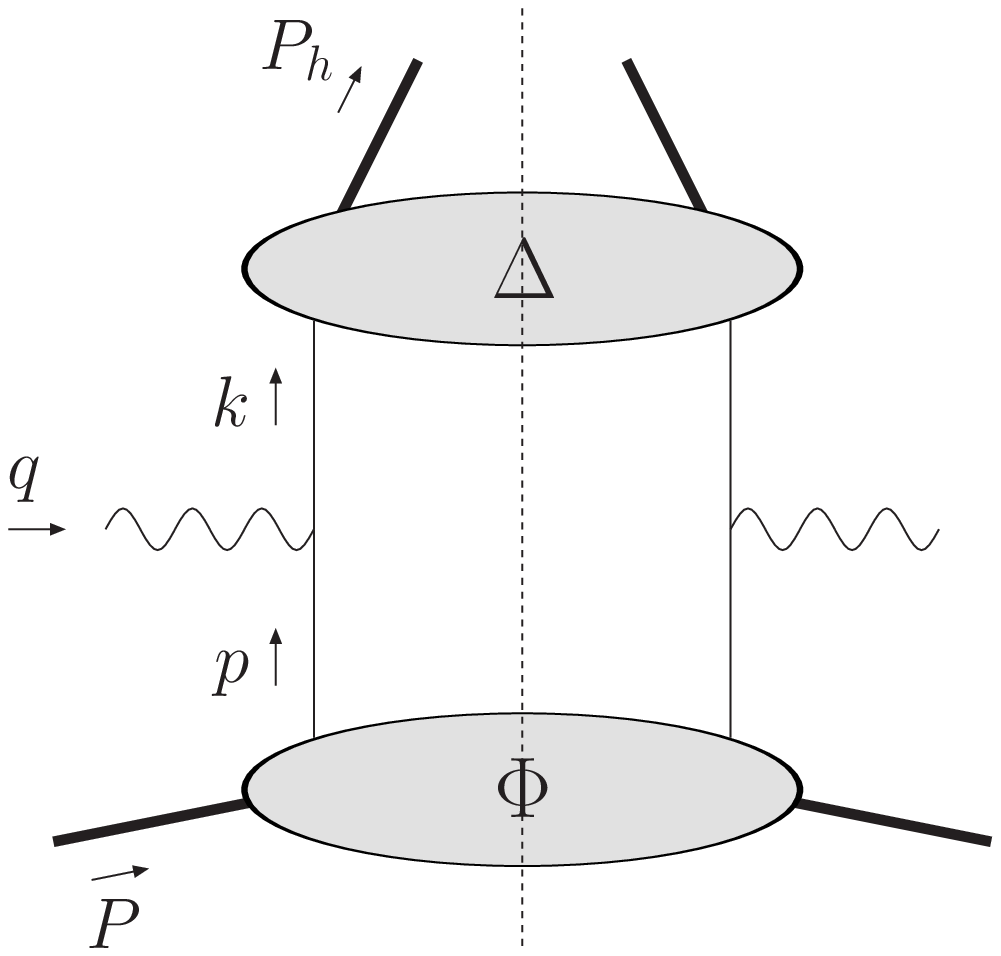}
&&
\includegraphics[height=3.5cm]{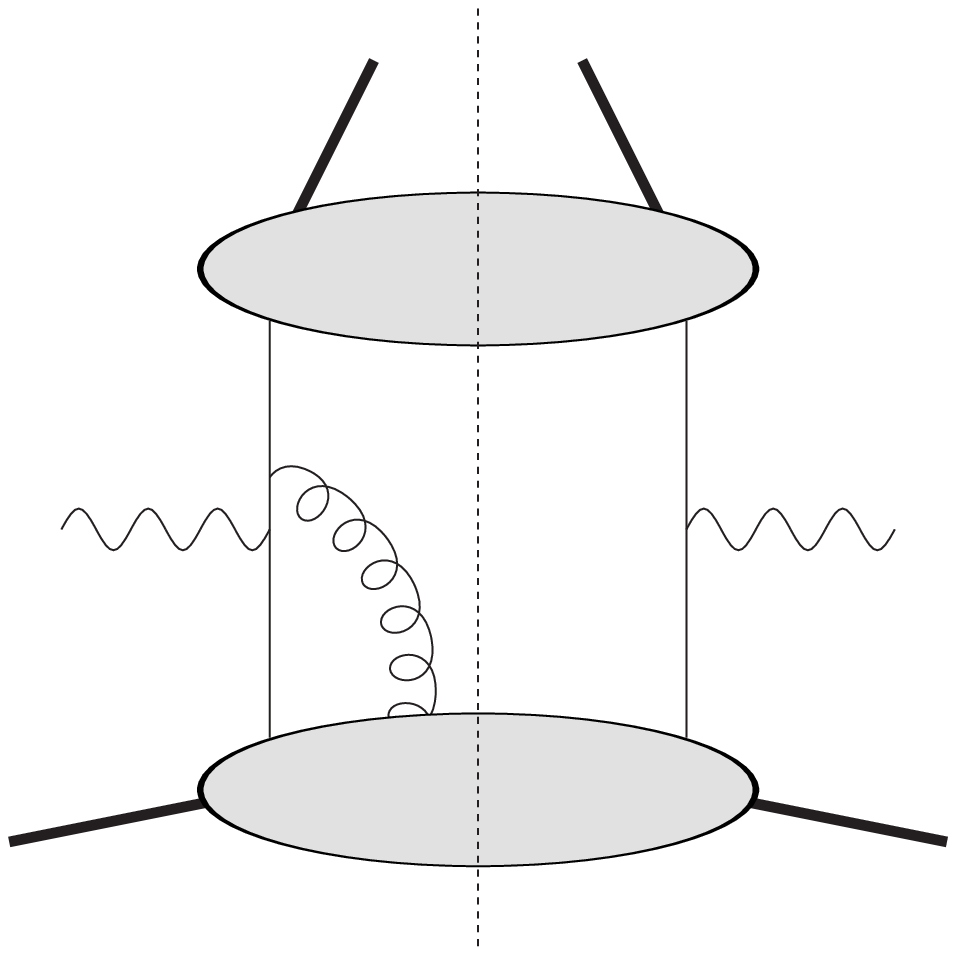}
&&
\includegraphics[height=3.5cm]{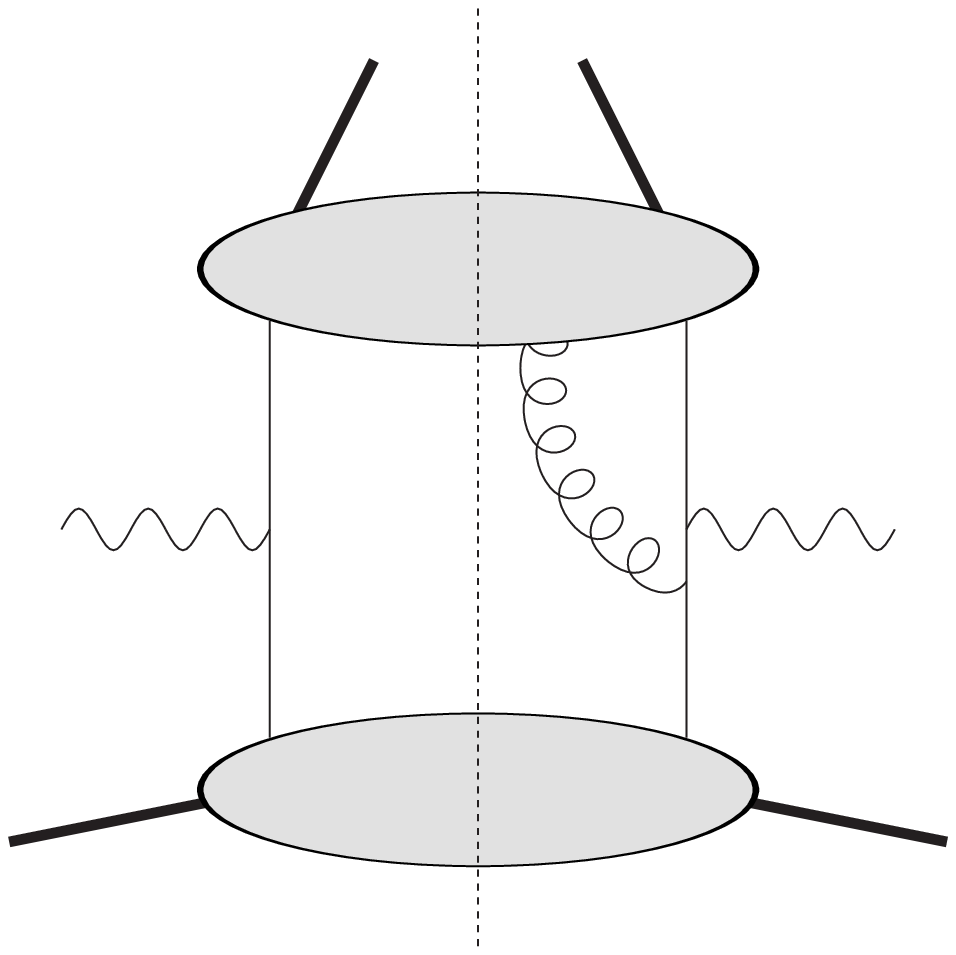}
\\
(a)
&&
(b)
&&
(c)
\end{tabular}
\caption{\label{f:graphs}  Examples of graphs
  contributing to semi-inclusive DIS at low transverse momentum of the
  produced hadron.}
%\end{center}
\end{figure}

In the kinematical limit where $Q^2$
becomes large while $x$, $z$ and $P_{h\perp}^2$ remain fixed, it
has been proven~\cite{Ji:2004wu} 
that the cross section up to leading order in $1/Q$
can be factorized into a hard 
photon-quark scattering process and {\em transverse momentum dependent
  distribution and fragmentation functions} (for the regime
$M^2 \ll P_{h\perp}^2 \ll Q^2$, see \cite{spin06highpt}).  
Here, it is  assumed that
factorization applies up to subleading order 
in the $1/Q$ expansion, and
only graphs with the hard scattering at tree level are considered. 
Loops can then only occur as
shown, e.g., in Fig.~\ref{f:graphs} b, c.
It is convenient to introduce the shorthand notations
%\vspace{-3pt}
\begin{align} 
\hat{\bm{h}}&=\bm{P}_{h \perp}/|\bm{P}_{h
  \perp}|
&
&{\rm and}
&
{\cal C}\bigl[ \ldots \bigr]
&= \xbj\,
\sum_a e_a^2 \int d^2 \bm{p}_T\,  d^2 \bm{k}_T^{}
\, \delta^{(2)}\bigl(\bm{p}_T - \bm{k}_T^{} - \bm{P}_{h \perp}/z \bigr)
\ldots,
%\vspace{-3pt}
\end{align} 
where the sum runs over the quark
and antiquark flavors $a$, and $e_a$ denotes their fractional charge.
The corresponding
expression of the hadronic tensor is~\cite{Boer:2003cm}
%\begin{align}
%\vspace{-3pt}
\begin{equation} 
\begin{split} 
2M &W^{\mu\nu} 
%&
= \frac{2z}{x} 
%\sum_a   e_a^2 \; \int
%\de^2 \bm{p}_T\ \de^2 \bm{k}_T^{}\ 
%\delta^2(\bm{p}_T + \bm{q}_T - \bm{k}_T^{}) \;
{\cal C} \Bigl\{\Tr\Bigl[
  \Phi^a (x,p_T) \gamma^\mu \Delta^a (z,k_T) \gamma^\nu \Bigr]
%\nonumber 
%\\
%& \qquad 
- \frac{1}{Q\sqrt{2}}\, 
\Tr \Bigl[
  \gamma^\alpha \gamma^- \gamma^\nu \slim
  \tilde{\Phi}^a_{A\slim \alpha}(x,p_T)\slim 
\\
&\quad\times
\gamma^\mu \Delta^a(z,k_T) 
%\\
+ \gamma^\alpha \gamma^+ \gamma^\mu
  \tilde{\Delta}^a_{A\slim \alpha} (z,k_T)\slim \gamma^\nu 
  \Phi^a (x,p_T)  
  + \mathrm{h.c.} \slim\Bigr] \Bigr\} + {\cal O}(1/Q^2).
\label{eq1}
\end{split} 
%\vspace{-3pt} 
\end{equation} 

The quark-quark correlators $\Phi$ and $\Delta$ 
can be decomposed up to order $1/Q$ 
in a general way in terms
of transverse-momentum-dependent distribution functions. 
The quark-gluon-quark correlators $\PhiA$ and $\DeltaA$ can be related to the
quark-quark correlators by means of the QCD equations of motion. Eventually,
this allows the computation of the hadronic tensor and of the structure
functions appearing in the cross section. 

The full calculation leads to the following results~\cite{Bacchetta:2006tn}
%\vspace{-3pt}
\begin{align} 
\label{F_UUT}
%\hspace{-3mm}
F_{UU ,T}
& 
= {\cal C}\bigl[ f_1 D_1 \bigr],_{\phantom{\bigr[}}
%\\
%\label{F_UUL}
\hspace{30mm}
F_{UU ,L} 
= 0,
\\ 
\label{F_UUcosphi}
%\hspace{-3mm}
F_{UU}^{\cos\phi_h}
& 
= \frac{2M}{Q}\,{\cal C}\biggl[
   - \frac{\hat{\bm{h}}\cdott \bm{k}_T^{}}{M_h}
   \biggl(\xbj  h\, H_{1}^{\perp } 
   + \frac{M_h}{M}\,f_1 \frac{\tilde{D}^{\perp }}{z}\biggr)
%\nonumber \\ & \qquad
   - \frac{\hat{\bm{h}}\cdott \bm{p}_T}{M}
     \biggl(\xbj f^{\perp } D_1
   + \frac{M_h}{M}\,h_{1}^{\perp } \frac{\tilde{H}}{z}\biggr)\biggr],
\\
\label{F_UUcos2phi}
%\hspace{-3mm}
F_{UU}^{\cos 2\phi_h}\!
& 
= {\cal C}\biggl[
   - \frac{2\, \bigl( \hat{\bm{h}}\cdott \bm{k}_T^{} \bigr)
   \,\bigl( \hat{\bm{h}}\cdott \bm{p}_T \bigr)
    -\bm{k}_T^{}\cdott \bm{p}_T}{M M_h}\,
    h_{1}^{\perp } H_{1}^{\perp }\biggr].
%\vspace{-3pt} 
\end{align} 
The distribution (fragmentation) functions $f_1$ ($D_1$), $f^{\perp}$
($\tilde{D}^{\perp}$), $h$ ($\tilde{H}$), and $h_1^{\perp}$ ($H_1^{\perp}$)
depend on
$x$ ($z$) and $p_T^2$ ($k_T^2$) and should carry a flavor index $a$. 

To order $1/Q$, the function $F_{UU,L}$ vanishes.
The
  structure function $F_{UU}^{\cos\phi_h}$ is subleading-twist, i.e.\ $1/Q$-suppressed compared to the
  other two nonvanishing ones. 
Measurements of $F_{UU}^{\cos\phi_h}$ 
have been reported in
  Refs.~\cite{Arneodo:1986cf, %Adams:1993hs,
Breitweg:2000qh}. %,Chekanov:2006gt}.
This structure function is associated with the
so-called Cahn effect.  If one neglects 
the contribution from interaction-dependent functions and T-odd
functions 
Eq.~(\ref{F_UUcosphi}) becomes
%\vspace{-3pt}
\begin{equation}
 F_{UU}^{\cos\phi_h}
\approx \frac{2M}{Q}\,{\cal C}\biggl[ -\frac{\hat{\bm{h}}\cdott
  \bm{p}_T}{M}\,f_1 D_1\biggr] .
%\vspace{-3pt}   
\end{equation} 
This coincides with the $\cos\phi_h$ term calculated to order $1/Q$ 
in the parton
model with intrinsic transverse momentum included in distribution
and fragmentation functions, see e.g.\ Eqs.~(32) and (33) in
 Ref.~\cite{Anselmino:2005nn}.

The structure function 
$F_{UU}^{\cos 2\phi_h}$ contains the functions $h_1^\perp$
  (Boer-Mulders function) and $H_1^\perp$ (Collins
  function). Measurements have been reported in
  Refs.~\cite{Breitweg:2000qh}.%,Chekanov:2006gt}.

Upon integration over the transverse momentum of the outgoing hadron, the
integrated semi-inclusive DIS result is recovered, i.e.,
%\vspace{-2pt}
\begin{align}
F_{UU ,T}(x,z,Q^2) &= \xbj\,\sum_a e_a^2\,f_1^a(\xbj)\,D_1^a(z),
&
F_{UU ,L}(x,z,Q^2) &= 0,
\end{align}
where
\vspace*{-24.5pt}
\begin{align}
f_1^a(x) &=  \int d^2 \bm{p}_T\; f_{1}^a(x,p_T^2),
&
 D_1^a (z) &= z^2 \int d^2 \bm{k}_T\,D_1^a (z, k_T^2).
%\vspace{-3pt} 
\end{align}
Finally, the standard inclusive DIS structure function is obtained by
%\vspace{-3pt}
\begin{equation} 
F_T (x,Q^2) =\int d z\ z\; F_{UU ,T}(x,z,Q^2) = 
\xbj\,\sum_a e_a^2\;f_1^a(\xbj).
%\vspace{-3pt}   
\end{equation} 

The analysis of the cross section and of the structure functions for
polarized beams and targets 
can be carried out in an analogous way. There are in total 18 structure
functions, eight are ${\cal O}(1/Q^0)$, eight are ${\cal O}(1/Q)$, and 
two are of
higher order.  One subleading-twist structure
function, $F_{LU}^{\sin\phi_h}$, 
appears if the beam is longitudinally polarized and the target
unpolarized. The only available measurement is by the CLAS
collaboration~\cite{Avakian:2003pk}.  Two 
structure functions, 
$F_{UL}^{\sin\phi_h}$ and $F_{UL}^{\sin 2\phi_h}$, 
appear if the beam is unpolarized and the
target longitudinally polarized. They have been measured only by the HERMES
collaboration~\cite{Airapetian:2005jc}. %,Airapetian:2002mf 
Two 
structure functions, $F_{LL}$ and
$F_{LL}^{\cos \phi_h}$, require longitudinal polarization of both beam and
target (see \cite{Kotzinian:2006ar} for a discussion of these terms).

The structure functions with transversely polarized targets are nine in
total. 
$F_{UT ,T}^{\sin(\phi_h -\phi_S)}$ and  $F_{UT}^{\sin(\phi_h +\phi_S)}$ 
received a lot of attention from the theoretical and
experimental side in the last years. They contain the Sivers and transversity
distribution functions, respectively. They have been discussed in several
talks during this conference~\cite{spin06}.

Among the remaining structure functions, I highlight here two illustrative
examples 
that have not been measured so far. The first one is
%\vspace{-3pt}
\begin{equation} 
F_{LT}^{\cos(\phi_h -\phi_S)}
 ={\cal C}\biggl[ \frac{\hat{\bm{h}}\cdott\bm{p}_T}{M} g_{1T}
D_1 \biggr] ,
%\vspace{-3pt}   
\end{equation} 
requiring both longitudinally polarized beam and transversely polarized
target. Through the measurement of this asymmetry and with the knowledge of the
unpolarized fragmentation functions, it is possible to measure the parton
distribution function $g_{1T}^q$, which is the only chiral-even, T-even,
leading-twist function in addition to the well-known unpolarized distribution
function, $f_1^q$, and helicity distribution function, $g_1^q$ (see also  \cite{Kotzinian:2006ar}). 

Another unmeasured structure function is  
%\vspace{-3pt}
\begin{equation} 
\begin{split} 
\label{e:FUTsinphiS}
F_{UT}^{\sin \phi_S }
&
 = \frac{2M}{Q}\,{\cal C}\biggl\{
   \biggl(\xbj  f_T   D_1
   - \frac{M_h}{M} \, h_{1}  \frac{\tilde{H}}{z}\biggr) 
%\nonumber 
\\ & \qquad
%\hspace{-9mm}
   - \frac{\bm{k}_T^{}\cdott \bm{p}_T}{2 M M_h}\,
     \biggl[\biggl(\xbj  h_{T}  H_{1}^{\perp } 
   + \frac{M_h}{M} g_{1T} \,\frac{\tilde{G}^{\perp }}{z}\biggr)
%\nonumber \\ & \qquad
   -  \biggl(\xbj  h_{T}^{\perp }  H_{1}^{\perp } 
   - \frac{M_h}{M} f_{1T}^{\perp } \,\frac{\tilde{D}^{\perp }}{z}
   \biggr) \biggr]\biggr\},
\end{split} 
%\vspace{-3pt}   
\end{equation} 
generating a $\sin\phi_S$ modulation in single-spin
asymmetries with a transversely polarized target. 
%What makes this structure
%function special is that 
Upon integration over the transverse momentum of the
outgoing hadron, only one term survives, namely
%\vspace{-3pt}
\begin{equation}
F_{UT}^{\sin \phi_S}(x,z,Q^2)=-\xbj\,\sum_a e_a^2\,
        \frac{2M_h}{Q}\,h_{1}^a(\xbj)\,\frac{\tilde{H}^a(z)}{z},
\label{e:sinphiSint}
%\vspace{-3pt}   
\end{equation} 
where the transversity distribution function appears, multiplied by a
subleading-twist fragmentation function. It could therefore be a good
observable for transversity studies.

Another interesting limit of Eq.~\eqref{e:FUTsinphiS} is that of
semi-inclusive jet production, $ \ell(l) + N(P) \to \ell(l-q) +
\text{jet}(P_j) + X$.  The structure function for this
process can be obtained from  Eq.~\eqref{e:FUTsinphiS}
 by replacing
$D_1(z,k_T^2)$ with $\delta(1-z)\slim \delta^{(2)}(\bm{k}_T)$, setting
all other 
fragmentation functions to zero and integrating over $z$. This gives
%\vspace{-3pt}
\begin{equation}
F_{UT}^{\sin \phi_S}(x,P_{j\perp}^2,Q^2) = \xbj\,\sum_a e_a^2\,
        \frac{2M}{Q}\,\xbj  f_{T}^{a}(x,P_{j\perp}^2),
\label{e:sinphisjet} 
%\vspace{-3pt}   
\end{equation} 

In totally inclusive DIS
this structure function has to vanish due to time-reversal
invariance. Starting from Eq.~\eqref{e:sinphiSint} 
or Eq.~\eqref{e:sinphisjet} and performing the
required integrations, the following two relations can be derived
%\vspace{-3pt}
\begin{align}
 \sum_h \int d z \;\tilde{H}^a(z) &=0,
&
\int d^2 \bm{p}_T\; f_T(x,p_T^2)&=0.
%\vspace{-3pt} 
\end{align} 
The
second relation can also be used to establish a connection between the Sivers
function and the Qiu-Sterman effect (see, e.g.,
\cite{Bacchetta:2005xk}). 
%So far, the second relation has been proposed only in its integrated form. 
%If $f_T(x,p_T^2)$ itself does not vanish,
%to make the structure function vanish in 
%inclusive DIS
%a complete integration over the transverse momentum of the hadron (or
%jet) is necessary, which
%is experimentally impossible. This could suggest that $f_T(x,p_T^2)$ is 
%identically zero.

In conclusion, I presented a selection of the results obtained in
\cite{Bacchetta:2006tn}, where the cross section for semi 
inclusive deep inelastic
scattering off a polarized nucleon for low transverse momentum of the detected
hadron was analyzed, completing the existing literature. 
I pointed out that there are in general
18 structure functions, nine of which have been already measured in some
experiments. More measurements will hopefully come in the future from HERMES,
COMPASS, and JLab. 

\

\end{document}